\documentclass[prc,twocolumn,nofootinbib,showpacs]{revtex4}
\usepackage{natbib}
\usepackage{amssymb,amsbsy,amsmath,amsfonts}
\usepackage{graphicx}
\usepackage{float}
\usepackage{rotating}

\begin{document}
\title {Testing the nature of the $\Lambda$(1520) in the $J/\psi\rightarrow\bar{\Lambda}K^-p$ and
$J/\psi\rightarrow\bar{\Lambda}\pi^+\pi^-\Lambda$ reactions}

\author{L. S. Geng$^1$, E. Oset$^1$, and B. S. Zou$^2$}
 \affiliation{ 
 $^1$Departamento de F\'{\i}sica Te\'orica and IFIC,
Centro Mixto Universidad de Valencia-CSIC,
Institutos de Investigaci\'on de Paterna, Aptdo. 22085, 46071 Valencia, Spain\\
 $^2$ Institute of High Energy Physics, Chinese Academy of Sciences, \\
 P.O.Box 918(4), Beijing 100049, China}

\begin{abstract}
We study the reactions $J/\psi\rightarrow\bar{\Lambda}K^-p$ and
$J/\psi\rightarrow\bar{\Lambda}\pi^+\pi^-\Lambda$ with a unitary chiral approach. We predict
the ratio of the invariant mass distributions of these two reactions at the $\Lambda(1520)$ peak position,
which is free of the unknown production mechanism and reflects only the properties of the $\Lambda(1520)$.
An experimental measurement of this ratio will provide the couplings of the $\Lambda(1520)$ to its
decay channels, allowing to test the predictions of the chiral unitary approach on this
resonance, which appears as dynamically generated in that approach.

\end{abstract}
\date{\today}
 \maketitle

\section{Introduction}
The applications of unitary chiral approaches (U$\chi$PT) in studies of
low-lying baryons (mesons) have revealed many interesting aspects of these 
hadronic systems. One of these concerns the $\Lambda(1405)$, which is found to correspond to 
two poles on the complex plane, one
narrower around 1420 MeV and one broader around 1390 
MeV~\cite{Oller:2000fj,Hyodo:2002pk,GarciaRecio:2002td,Jido:2003cb,Borasoy:2005ie,Oller:2005ig,Oller:2006jw,Borasoy:2006sr}. This could have tremendous
effects on our understanding of the kaon-nucleon interaction, and furthermore, 
on the idea of superstrong kaon-nucleon interaction. The reason is quite simple 
because the 1420 MeV pole is
just 12 MeV below the kaon-proton threshold while the position of the 
nominal $\Lambda(1405)$ is another 15 MeV lower. The attraction for the nominal $\Lambda(1405)$, 
therefore, is much stronger than for the former.

The $\Lambda(1520)$ is not less
interesting. Studies based on unitary chiral approaches 
showed that the $\pi\Sigma^*$ channel plays such an important role
in the formation of this resonance that it qualifies as a bound state of the $\pi\Sigma^*$~\cite{Kolomeitsev:2003kt,Sarkar:2004jh,Sarkar:2005ap,Roca:2006sz}.
More specifically, the coupling of the $\Lambda(1520)$ to the $\pi\Sigma^*$ channel 
is almost  two times larger than its couplings to $\bar{K}N$ and $\pi\Sigma$, although 
the branching ratios of the $\Lambda(1520)$ to these two latter channels are about $45\%$ and $42\%$,
respectively, while the branching ratio to the former is only $\sim10\%$~\cite{Roca:2006sz}.\footnote{The
couplings of the $\Lambda(1520)$ to the various channels have been defined
in a way that the $d$-wave couplings incorporate the $q^2$ factor
and a normalization such that, up to phase space, the decay widths have the same expression
in terms of the coupling constants. Hence, 
a direct comparison of the $s$-wave and $d$-wave couplings is appropriate~\cite{Sarkar:2005ap,Roca:2006sz}.} The U$\chi$PT description of the
$\Lambda(1520)$ has recently been studied in various reactions, including
$K^-p\rightarrow \pi^+\pi^-\Lambda$, $\gamma p\rightarrow K^+ K^- p$, $\gamma p\rightarrow K^+ \pi^0\pi^0\Lambda$,  $\pi^- p\rightarrow K^0 K^- p$~\cite{Roca:2006sz}, $K^-p\rightarrow\pi^0\pi^0\Lambda$~\cite{Sarkar:2005ap,Roca:2006sz}, 
$pp\rightarrow p K^+ K^- p$, $pp\rightarrow p K^+\pi^0\pi^0\Lambda$~\cite{Roca:2006pu},
and the radiative decay of the $\Lambda(1520)$ to  $\gamma\Lambda$ ($\gamma\Sigma^0$)~\cite{Doring:2006ub}.

The $J/\psi$ and $\psi'$ experiments at the Beijing Electron-Positron Collider (BEPC)
provide an excellent place for studying excited nucleons and baryons, due to the
spin- and isospin-filtering of the $J/\psi$ decay process~\cite{Zou:2000wg,Ablikim:2004ug}.
The BEPC is now being upgraded to a two-ring collider
(BEPCII) with a design luminosity of $1\times10^{33}$ cm$^{−2}$ s$^{−1}$ at 3.89 GeV and will
operate between 2 and 4.2 GeV in the center of mass. With this luminosity,
the new BESIII detector will be able to collect, for example, 10 billion $J/\psi$
events in one year of running. This amount of $J/\psi$ events will make possible a 
complementary study of the properties of many excited nucleons and hyperons, which
in the past were only accessible in $\pi N$ ($KN$) scatterings. In this sense, this work
also aims to motivate the BES collaboration to perform a detailed study of the
proposed reactions.

This paper proceeds as follows: In section II, we briefly outline the
unitary chiral description of the $\Lambda(1520)$. In Section III, we 
study the $J/\psi\rightarrow \bar{\Lambda}K^-p$ and 
$J/\psi\rightarrow\bar{\Lambda} \pi^+\pi^-\Lambda$ reactions 
within the unitary chiral approach, more specifically, we look at the ratio of
the invariant mass distributions at the $\Lambda(1520)$ peak position.
By studying this quantity, we not only can deepen our understanding of the nature 
of the $\Lambda(1520)$ and the underlying chiral dynamics, but also can avoid many
uncertainties related to the unknown vertices and couplings in the production of
the $\Lambda(1520)$. Results and discussions are given in Section IV, followed by a brief summary 
in Section V.

\begin{table*}[htpb]
      \renewcommand{\arraystretch}{1.5}
     \setlength{\tabcolsep}{0.4cm}
     \centering
     \caption{The parameter values in the dynamical generation of the
$\Lambda(1520)$. The parameters $\gamma_{13}$ and $\gamma_{14}$ are in units of $10^{-7}$ MeV$^{-5}$, while
$\gamma_{33}$, $\gamma_{34}$ and $\gamma_{44}$ are in $10^{-12}$ MeV$^{-5}$.  \label{table:cpc1}}
     \begin{tabular}{ccccc}
     \hline\hline 
       $\gamma_{13}$  &$\gamma_{14}$  &$\gamma_{33}$  &
$\gamma_{34}$  &$\gamma_{44}$  \\
\hline
 $0.98\pm0.04$ & $1.10\pm0.04$ & $-1.73\pm0.02$ & $-1.108\pm0.010$ &
$-0.730\pm0.016$\\
    \hline\hline
    \end{tabular} 
       \end{table*}
\section{The $\Lambda(1520)$ as a dynamically generated resonance}
The $\Lambda(1520)$ is dynamically generated in a
coupled channel formalism, including the
$\pi\Sigma^*$, $K\Xi^*$ in $s$ waves and $\bar{K}N$ and
$\pi\Sigma$ in $d$ waves~\cite{Sarkar:2005ap,Roca:2006sz}.
The inclusion of the $d-$wave coupled channels is essential to
put the $\Lambda(1520)$ at the right position with a proper width.

The matrix containing the tree-level amplitudes is written as (in the
order of $\pi\Sigma^*$, $K\Xi^*$, $\bar{K}N$, and $\pi\Sigma$)~\cite{Roca:2006sz}:
\begin{equation}\label{eq:treeV}
 V=\left(\begin{array}{cccc}
C_{11}(k_1^0+k_1^0)&C_{12}(k_1^0+k_2^0)&\gamma_{13}q^2_3&\gamma_{14}q^2_4\\
C_{21}(k_2^0+k_1^0)&C_{22}(k_2^0+k_2^0)&0&0\\
\gamma_{13}q^2_3&0&\gamma_{33}q^4_3&\gamma_{34}q^2_3q^2_4\\
\gamma_{14}q^2_4&0&\gamma_{34}q^2_3q^2_4&\gamma_{44}q^4_4\\
         \end{array}\right),
\end{equation}
where $q_i=\frac{\sqrt{(s-(M_i+m_i)^2)(s-(M_i-m_i)^2)}}{2\sqrt{s}}$ 
is the center-of-mass momentum of channel $i$, $k^0_i=\frac{s+m_i^2-M_i^2}{2\sqrt{s}}$ is the energy
of meson $i$ with $M_i$ ($m_i$) the baryon (meson) mass of channel $i$. The coefficients $C_{ij}$ are
provided by the lowest-order chiral Lagrangian describing the interaction of the
decuplet of the $\Delta(1232)$ and the octet of the $\pi$~\cite{Kolomeitsev:2003kt,
Sarkar:2004jh} and have the following values
\begin{equation}
 C_{11}=-\frac{1}{f^2},\quad C_{12}=C_{21}=\frac{\sqrt{6}}{4 f^2},\quad C_{22}=-\frac{3}{4f^2}.
\end{equation}
In Ref.~\cite{Roca:2006sz}, $f=1.15f_\pi$ is used 
with $f_\pi=93$ MeV, the pion decay constant. In the present work, we also use the same value.
The $K\Xi^*$ coupling to the $\bar{K}N$ and $\pi\Sigma$ channels through $d$ wave interactions
have been neglected since the $K\Xi^*$ threshold is far from the $\Lambda(1520)$ and, therefore, plays
a minor role in the description of this resonance.

The $\gamma_{ij}$'s are not determined by chiral dynamics and their values (together
with the subtraction constants appearing in the loop calculations, see below) are fitted
to the experimental results on the $\bar{K}N$ and $\pi\Sigma$ partial-wave amplitudes~\cite{Roca:2006sz}. The explicit values are given in
Table \ref{table:cpc1}.

Several unitarization procedures have been developed over the years to
unitarize the coupled channel amplitudes, including
the inverse amplitude method~\cite{Dobado:1996ps}, the N/D method~\cite{Oller:1998zr}, 
and the Bethe-Salpeter approach~\cite{Oller:1997ti}. They differ from
the inclusion of higher orders, a left cut, or tadpole and
crossed channels. These differences may be relevant at very low
energies but at higher energies they have been shown to be minute when resonances
are already dynamically generated using the lowest order interaction kernels.

In the present work, following Refs.~\cite{Sarkar:2005ap,Roca:2006sz}, we 
unitarize the kernel $V$ of Eq.~(\ref{eq:treeV}) through the Bethe-Salpeter approach
to obtain the unitarized amplitude $T$:
\begin{equation}
 T=V+VGT=(1-VG)^{-1}V,
\end{equation}
where $G$ is a diagonal matrix of the loop functions of one-baryon and one-meson
\begin{eqnarray}
 G_j&=&i 2M_j\int\frac{d^4q}{(2\pi)^4}\frac{1}{(P-q)^2-M_j^2+i\epsilon}\frac{1}{q^2-m_j^2+i\epsilon}\nonumber\\
&=&\frac{2M_j}{16\pi^2}\Bigg\{a_j(\mu)+\log\frac{M_j^2}{\mu^2}
+\frac{s+m_j^2-M_j^2}{2s}\log\frac{m^2_j}{M_j^2}\nonumber\\
&&\hspace{1.0cm}\times\frac{q}{\sqrt{s}}\bigg[\log(s^{-+}_j s^{++}_j)-\log(s^{--}_j s^{+-}_j)\bigg]\Bigg\},
\end{eqnarray}
\begin{equation}
s^{\pm\pm}_j=s\pm(M_j^2-m_j^2)\pm 2 q\sqrt{s},
\end{equation}
where $\mu$ is the scale of the dimensional regularization (in the present work,
we use $\mu=700$ MeV), $s=P^2$ and $P$ is the total four momentum of the
meson-baryon system, which is $(\sqrt{s},0,0,0)$ in the rest frame of the meson-baryon system.
The subtraction constants are fitted to the  experimental results on the $\bar{K}N$ and $\pi\Sigma$ partial-wave
amplitudes together with the $\gamma_{ij}$'s as mentioned above to yield $a_s=-1.78\pm0.02$ and $a_d=-8.13\pm0.03$ for $\pi\Sigma^*
$ ($K\Xi^*$) and $\bar{K}N$ ($\pi\Sigma$) channels, respectively~\cite{Roca:2006sz}. It is worthwhile 
mentioning that the $\pi\Sigma^*$ amplitudes are not included in the fit and, therefore, are 
the predictions of the model.

\begin{figure}[htpb]
\includegraphics[scale=0.35,angle=270]{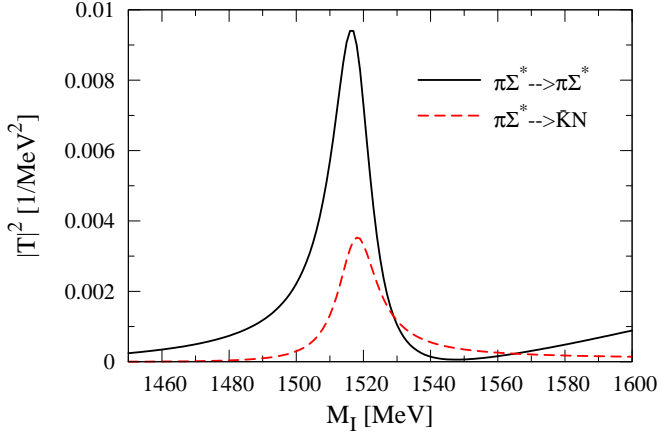}
\caption{Moduli squared of the amplitudes 
$T_{\pi\Sigma^*\rightarrow\pi\Sigma^*}$ and $T_{\pi\Sigma^*\rightarrow \bar{K}N}$.\label{fig:uchptA}}
\end{figure}

The moduli squared of the U$\chi$PT amplitudes are shown in Fig.~\ref{fig:uchptA}, where
the importance of the $\pi\Sigma^*$ channel can be clearly seen. The ratio
of $\frac{|T_{\pi\Sigma^*}|^2}{|T_{\bar{K}N}|^2}$ at the peak position is $\sim2.7$.

The above scattering matrix $T$ obtained in the unitary framework is related to the
$t$ matrix with the Mandl-Shaw normalization in the following way
~\cite{Sarkar:2005ap,Roca:2006sz} (we only list those
needed later):
\begin{equation}
 t_{\pi\Sigma^*\rightarrow\pi\Sigma^*}=T_{\pi\Sigma^*\rightarrow\pi\Sigma^*},
\end{equation}
\begin{eqnarray}
 t_{\bar{K}N\rightarrow\pi\Sigma^*}&=&T_{\bar{K}N\rightarrow\pi\Sigma^*}C\left(\frac{1}{2}2\frac{3}{2};m,M-m\right)\nonumber\\
 &&\times Y_{2,m-M}(\hat{k})(-1)^{M-m}\sqrt{4\pi},
\end{eqnarray}
where $\hat{k}$ is the unit vector of the kaon momentum in the $\bar{K}N$ center of mass system.

\section{Reaction mechanisms of $J/\psi\rightarrow\bar{\Lambda}K^-p$ and $J/\psi\rightarrow\bar{\Lambda}\pi^+\pi^-\Lambda$}
\begin{figure}[htpb]
\includegraphics[scale=0.5]{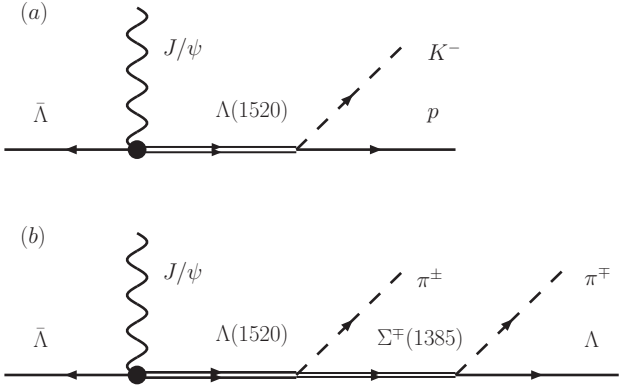}
\caption{The reaction mechanisms of $J/\psi\rightarrow \bar{\Lambda} K^- p$ and $J/\psi\rightarrow \bar{\Lambda}\pi^+\pi^-\Lambda$.\label{fig:diagram}}
\end{figure}
\begin{figure*}[htpb]
\includegraphics[scale=0.8]{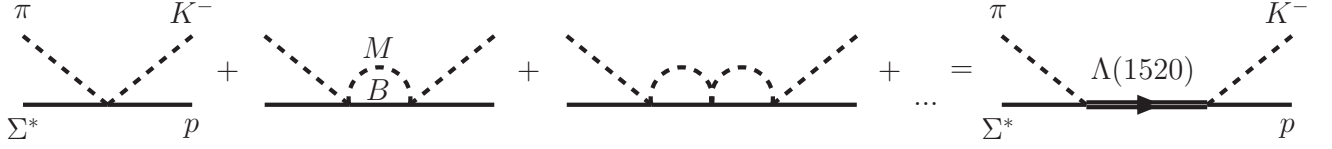}
\caption{A schematic representation of the $\Lambda(1520)$ as dynamically
generated in the $\pi\Sigma^*\rightarrow K^- p$ interaction.\label{fig:diagram2}}
\end{figure*}
The dynamical generation of the $\Lambda(1520)$ implies that
the reactions proceed in the following way:  the $J/\psi$ decays into $\bar{\Lambda}MB$; the rescattering
of the meson baryon, $MB$, pairs generates dynamically the $\Lambda(1520)$, which then decays into either $K^-p$ or $\pi\Sigma(1385)$. In the latter
case, the $\Sigma(1385)$ propagates and decays into  $\pi\Lambda$. Such
a process is illustrated in Fig.~\ref{fig:diagram}.

Since we are only interested in the
ratio of the corresponding invariant mass distributions of these two reaction around the $\Lambda(1520)$ peak position, the vertex of the $J/\psi$
decaying into $\bar{\Lambda}\Lambda(1520)$ is not relevant because it appears in both reactions and
cancels in the ratio. Thus, this removes the uncertainties of our study related to this unknown vertex.

For the $J/\psi\rightarrow \bar{\Lambda} K^- p$ reaction, since we will
integrate over the whole phase space of the $K^-p$ particles, the frame of reference is
irrelevant. There will always be  a frame of reference where the $\Lambda(1520)$ has
the polarization of $+3/2$. Hence we shall study the decay of the $\Lambda(1520)$ with $S_z=3/2$.
The amplitude of Fig.~\ref{fig:diagram}a contains the unknown $\bar{\Lambda} J/\Psi\Lambda(1520)$
vertex, the $\Lambda(1520)$ propagator and its coupling to $K^-p$. This amplitude is
proportional to $t_{\pi\Sigma^*\rightarrow K^-p}$, which contains implicitly the
$\Lambda(1520)$ propagator (the $\Lambda(1520)$ appears as a pole of this amplitude) and the coupling of the resonance to $K^- p$, as depicted in Fig.~\ref{fig:diagram2}. Hence, 

\begin{equation}
t_1\sim  t_{\pi\Sigma^*\rightarrow K^- p}=\frac{1}{\sqrt{2}}t_{\pi\Sigma^*\rightarrow \bar{K}N},
\end{equation}
where $\frac{1}{\sqrt{2}}$ accounts for the projection of the isospin 0 state to the charged state.
In the above amplitude, we have neglected all the factors common to this reaction and the 
$J/\psi\rightarrow\bar{\Lambda}\pi^+\pi^-\Lambda$ reaction. The modulus squared of the
amplitude is
\begin{eqnarray}
|t_1|^2&=&\frac{4\pi}{2}|T_{\pi\Sigma^*\rightarrow\bar{K}N}|^2\times\nonumber\\
&&\sum_{m=\frac{1}{2},-\frac{1}{2}}C(\frac{1}{2}2 \frac{3}{2};m,\frac{3}{2}-m)^2 |Y_{2,m-\frac{3}{2}}(\vec{k})|^2 \nonumber\\
&=&\frac{3}{4}\sin^2(\theta) |T_{\pi\Sigma^*\rightarrow\bar{K}N}|^2.
\end{eqnarray}
Since we will integrate later over phase space, we replace $\frac{3}{4}\sin^2(\theta)$ by
\begin{equation}
\frac{\int d\phi\int^1_{-1}d\cos(\theta)\frac{3}{4}\sin^2(\theta)}{\int d\phi\int^{1}_{-1}d\cos(\theta)}=\frac{1}{2}.
\end{equation}
Therefore, one can effectively substitute 
\begin{equation}
|t_1|^2\rightarrow\frac{1}{2}|T_{\pi\Sigma^*\rightarrow\bar{K}N}|^2.
\end{equation}

For the $J/\psi\rightarrow\bar{\Lambda}\pi^+\pi^-\Lambda$ reaction, the corresponding $t$ matrix element
reads
\begin{equation}
-it_2=
-\frac{1}{\sqrt{3}}T_{\pi\Sigma^*\rightarrow\pi\Sigma^*}\frac{f_{\pi\Sigma^*\Lambda}}{m_\pi}S_j\left[\tilde{q}^+_j \Pi_{\Sigma^{*+}}+\tilde{q}^-_j \Pi_{\Sigma^{*-}}
\right],
\end{equation}
\begin{equation}
\Pi_{\Sigma^{*\pm}}=\frac{1}{\sqrt{s'_{\Sigma^{*\pm}}}-M_{\Sigma^*}+\frac{i}{2}\Gamma(s'_{\Sigma^{*\pm}})},
\end{equation}
where the factor $\frac{1}{\sqrt{3}}$ again accounts for the projection of
the isospin state to the physical state, $\vec{S}$ is the spin 1/2 to 3/2 transition operator, and $\tilde{q}^+$ ($\tilde{q}^-$) is the momentum of
$\pi^+$ ($\pi^-$) in the rest frame of $\Sigma^{*+}$ ($\Sigma^{*-}$). The coupling constant
$f_{\pi\Sigma^*\Lambda}=1.3$ is determined in Ref.~\cite{Sarkar:2005ap} to reproduce
the partial decay width of 32 MeV for $\Sigma^{*0}\rightarrow\pi^0\Lambda$. As in
Ref.~\cite{Roca:2006sz}, we have introduced an energy dependent width for the
$\Sigma^*$
\begin{eqnarray}
\Gamma(s')&=&\Gamma_0\bigg[0.88\frac{q^3(s',M_\Lambda^2,m^2_\pi)}{q^3(M_{\Sigma^*}^2,M_\Lambda^2,m^2_\pi)}
\Theta(\sqrt{s'}-M_\Lambda-m_\pi)\nonumber\\ 
&&\hspace{0.6cm}+0.12\frac{q^3(s',M_\Sigma^2,m^2_\pi)}{q^3(M_{\Sigma^*}^2,M_\Sigma^2,m^2_\pi)}
\Theta(\sqrt{s'}-M_\Sigma-m_\pi)\bigg],\nonumber
\end{eqnarray}
where $\Gamma_0=37.1$ MeV~\cite{Yao:2006px}.

Since the width of this channel
does not depend on the third component of the $\Sigma^*$, we can average over the $\Sigma^*$ spin and sum over
the $\Lambda$ spin. Using the following relation
\begin{eqnarray}
&&\frac{1}{4}\sum_{M=-3/2}^{3/2}\sum_{m=-1/2}^{1/2}\langle m|S_i|M\rangle\langle M|S^\dagger_j|m\rangle\nonumber\\
&&=\frac{1}{4}\sum_m\langle m|\frac{2}{3}\delta_{ij}-\frac{i}{3}\epsilon_{ijk}\sigma_k|m\rangle
=\frac{1}{3}\delta_{ij},
\end{eqnarray}
one obtains
\begin{eqnarray}
|t_2|^2&=&\frac{1}{9}\left(\frac{f_{\pi\Sigma^*\Lambda}}{m_\pi}\right)^2 |T_{\pi\Sigma^*\rightarrow\pi\Sigma^*}|^2\times\\
&&\bigg\{
\left|\Pi_{\Sigma^{*+}}\right|^2|\tilde{q}^+|^2
+\left|\Pi_{\Sigma^{*-}}\right|^2|\tilde{q}^-|^2\nonumber\\
&&\hspace{0.5cm}+2\tilde{q}^+\cdot\tilde{q}^-\mathrm{Re}
\left[\Pi_{\Sigma^{*+}}
\left(\Pi_{\Sigma^{*-}}\right)^*\right]\bigg\}.\nonumber
\end{eqnarray}


The total decay widths of these two reactions, with the amplitudes squared obtained above,
are 
\begin{eqnarray}
\Gamma(J/\psi\rightarrow\bar{\Lambda}K^-p)&=&\frac{2M_{\bar{\Lambda}} M_p}{M_{J/\psi}}\frac{1}{(2\pi)^5}\times\\
&&\hspace{-2.8cm}\int \frac{ d^3q_{\bar{\Lambda}}}{2E_{\bar{\Lambda}}}\int\frac{d^3q_p}{2E_p}\int\frac{d^3q_K}{2\omega_K} |t_1|^2
\delta^{4}(q_{\bar{\Lambda}}+q_{K^-}+q_p),\nonumber
\end{eqnarray}
\begin{eqnarray}
\Gamma(J/\psi\rightarrow\bar{\Lambda}\pi^+\pi^-\Lambda)&=&\frac{2M_{\bar{\Lambda}} M_\Lambda}{M_{J/\psi}}\frac{1}{(2\pi)^8}\times\\
&&\hspace{-2cm}\int \frac{ d^3q_{\bar{\Lambda}}}{2E_{\bar{\Lambda}}}\int\frac{d^3q_\Lambda}{2E_\Lambda}
\int\frac{d^3q_{\pi^+}}{2\omega_{\pi^+}}\int\frac{d^3q_{\pi^-}}{2\omega_{\pi^-}} |t_2|^2\nonumber\\
&&\hspace{-0.1cm}\times
\delta^4(q_{\bar{\Lambda}}+q_{\Lambda}+q_{\pi^+}+q_{\pi^-}),\nonumber
\end{eqnarray}
where $M_{\bar{\Lambda}}$ ($E_{\bar{\Lambda}}$), $M_\Lambda$ ($E_\Lambda$), 
and $M_p$ ($E_p$) are the masses (energies) of the $\bar{\Lambda}$, the $\Lambda$, and
the proton, while $\omega_K$, $\omega_{\pi^+}$ and $\omega_{\pi^-}$ are the energies
of the kaon, the $\pi^+$, and the $\pi^-$. The above integrals are calculated 
with Monte-Carlo method and events within a given invariant mass interval
are collected to yield the corresponding invariant mass distributions, $d\Gamma/d M_I$.

\begin{figure}[t]
\includegraphics[scale=0.35,angle=270]{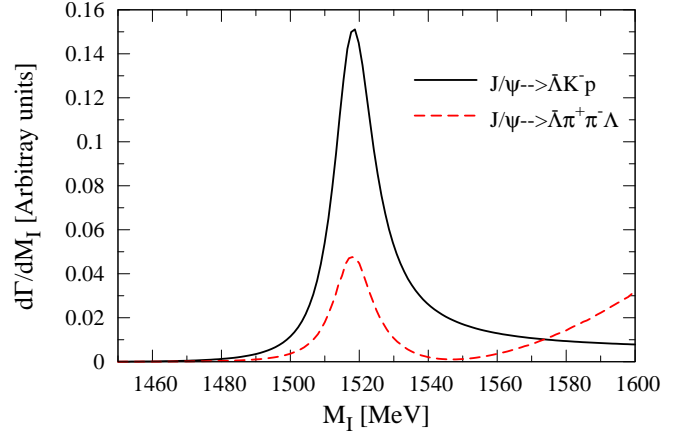}
\caption{The invariant mass distributions of the $J/\psi\rightarrow\bar{\Lambda}K^-p$ and
$J/\psi\rightarrow\bar{\Lambda}\pi^+\pi^-\Lambda$ reactions.\label{fig:ratio}}
\end{figure}

\section{Results and discussions}

We show in Fig.~\ref{fig:ratio} the invariant mass distributions for
the $J/\psi\rightarrow\bar{\Lambda}K^-p$ and $J/\psi\rightarrow \bar{\Lambda}\pi^+\pi^-\Lambda$
reactions, where the $\Lambda(1520)$ can be clearly seen. The ratio of the invariant mass distributions
at the peak position is 
\begin{equation}
\frac{d\sigma/d M_I(J/\psi\rightarrow\bar{\Lambda}K^-p)}{d\sigma/d M_I(J/\psi\rightarrow \bar{\Lambda}\pi^+\pi^-\Lambda)}\bigg|_{M_I\sim1520\,\mathrm{MeV}}\approx3.2.
\end{equation}
This translates into
\begin{equation}
\frac{d\sigma/d M_I(J/\psi\rightarrow\bar{\Lambda}\bar{K}N)}{d\sigma/d M_I(J/\psi\rightarrow \bar{\Lambda}\pi\pi\Lambda)}\bigg|_{M_I\sim1520\,\mathrm{MeV}}\approx4.3.
\end{equation}

\begin{figure}[t]
\includegraphics[scale=0.35,angle=270]{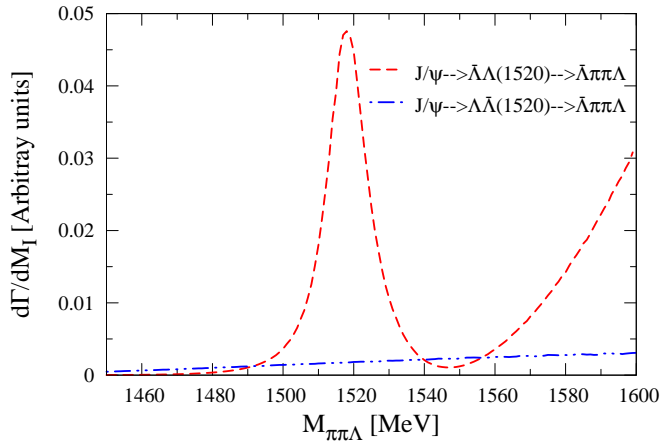}
\caption{The invariant mass distributions of the 
$J/\psi\rightarrow\bar{\Lambda}\Lambda(1520)\rightarrow\bar{\Lambda}\pi\pi\Lambda$ and 
$J/\psi\rightarrow\Lambda\bar{\Lambda}(1520)\rightarrow\bar{\Lambda}\pi\pi\Lambda$ reactions.\label{fig:lambdabar}}
\end{figure}

Since the $\Lambda(1520)$ is a rather narrow resonance with a width of 15.6 MeV, the above
ratio can be compared with the branching ratio of the $\Lambda(1520)$ decaying into 
$\bar{K}N$ and $\pi\pi\Lambda$. Using the numbers given in the PDG~\cite{Yao:2006px}, $\sim45\%$ for $\bar{K}N$ and
$\sim10\%$ for $\pi\pi\Lambda$, we obtain $R\sim4.5$ and indeed it is quite close to our predicted number.
However, it should be noted that the PDG branching ratio for the $\Lambda(1520)$  decaying into 
$\pi\pi\Lambda$ include various mechanisms, where the $\Lambda(1520)$ does not 
necessarily decay first into $\pi\Sigma^*$ with
subsequent  $\Sigma^*$ decay into $\pi\Lambda$. In fact three experiments are quoted in
the PDG claiming
$\Gamma(\Sigma^*\pi\rightarrow\Lambda\pi\pi)/\Gamma(\Lambda\pi\pi)$ equal to
either $0.39\pm0.10$~\cite{Burkhardt:1971tb},  $0.82\pm0.10$~\cite{Mast:1973gb}, or $0.58\pm0.22$~\cite{Corden:1975},

In the formalism followed here we have created $\bar{\Lambda}\Lambda(1520)$.
However, we could as well produce $\Lambda\bar{\Lambda}(1520)$. In this latter case
the $\bar{\Lambda}$(1520) would decay into $K^+\bar{p}$, a channel we are not concerned about, but
also into $\pi\pi\bar{\Lambda}$ and we would finish having the same final state $\Lambda\pi\pi\bar{\Lambda}$
as in the case of $\bar{\Lambda}\Lambda(1520)$ production. Since these two reactions 
have the same final state, their amplitudes would sum. Yet, it is easy to see that
this new mechanism has a negligible effect on the calculated $d\Gamma/dM_I$ around
the peak of the $\Lambda(1520)$. Indeed, $d\Gamma/dM_I(\pi\pi\Lambda)$ from
$J/\Psi\rightarrow\Lambda\bar{\Lambda}(1520)$ can be obtained from 
$d\Gamma/dM_I(\pi\pi\bar{\Lambda})$ of the $J/\Psi\rightarrow\bar{\Lambda}\Lambda(1520)$
reaction. In Fig.~\ref{fig:lambdabar} we show these two invariant mass distributions and we see that the one coming from $J/\Psi\rightarrow\Lambda\bar{\Lambda}(1520)$ below the peak of the
$\Lambda(1520)$ is extremely small and cannot change the results already obtained. The
background below the peak of $d\Gamma/dM_I$ is of the order of $3\%$. Even if one had
a coherent sum of the amplitudes, the effect of this background would still be moderate.
However, since in $d\Gamma/dM_I$ for  a given $M_I$ there are contributions of different pion momenta
because of the four body phase space, and also there are contributions of different
spins, the two mechanisms are highly incoherent and thus, the background from $J/\Psi\rightarrow
\Lambda\bar{\Lambda}(1520)$ can be safely ignored in the analysis.
\begin{figure*}[htpb]
\includegraphics[scale=0.35,angle=270]{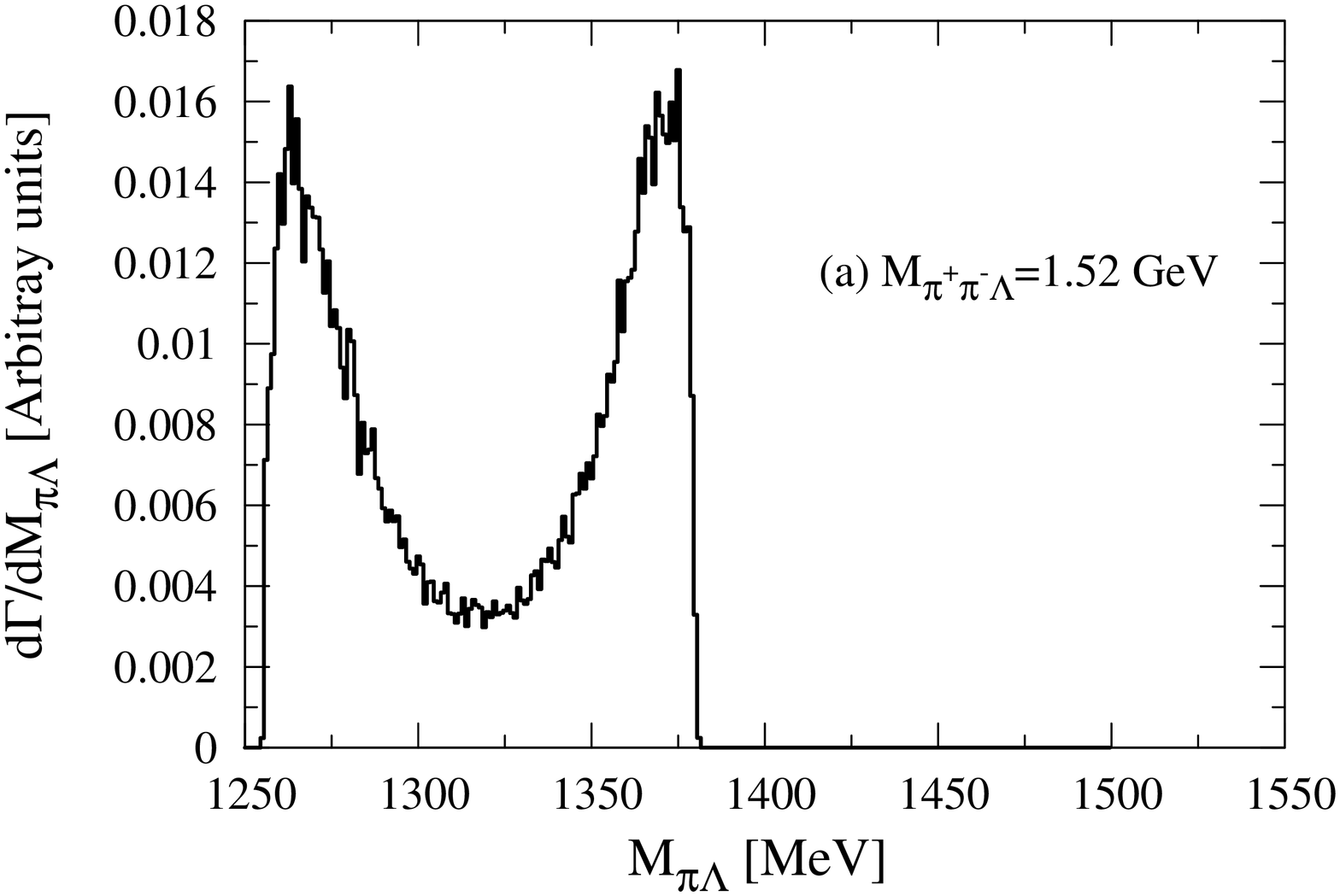}
\includegraphics[scale=0.35,angle=270]{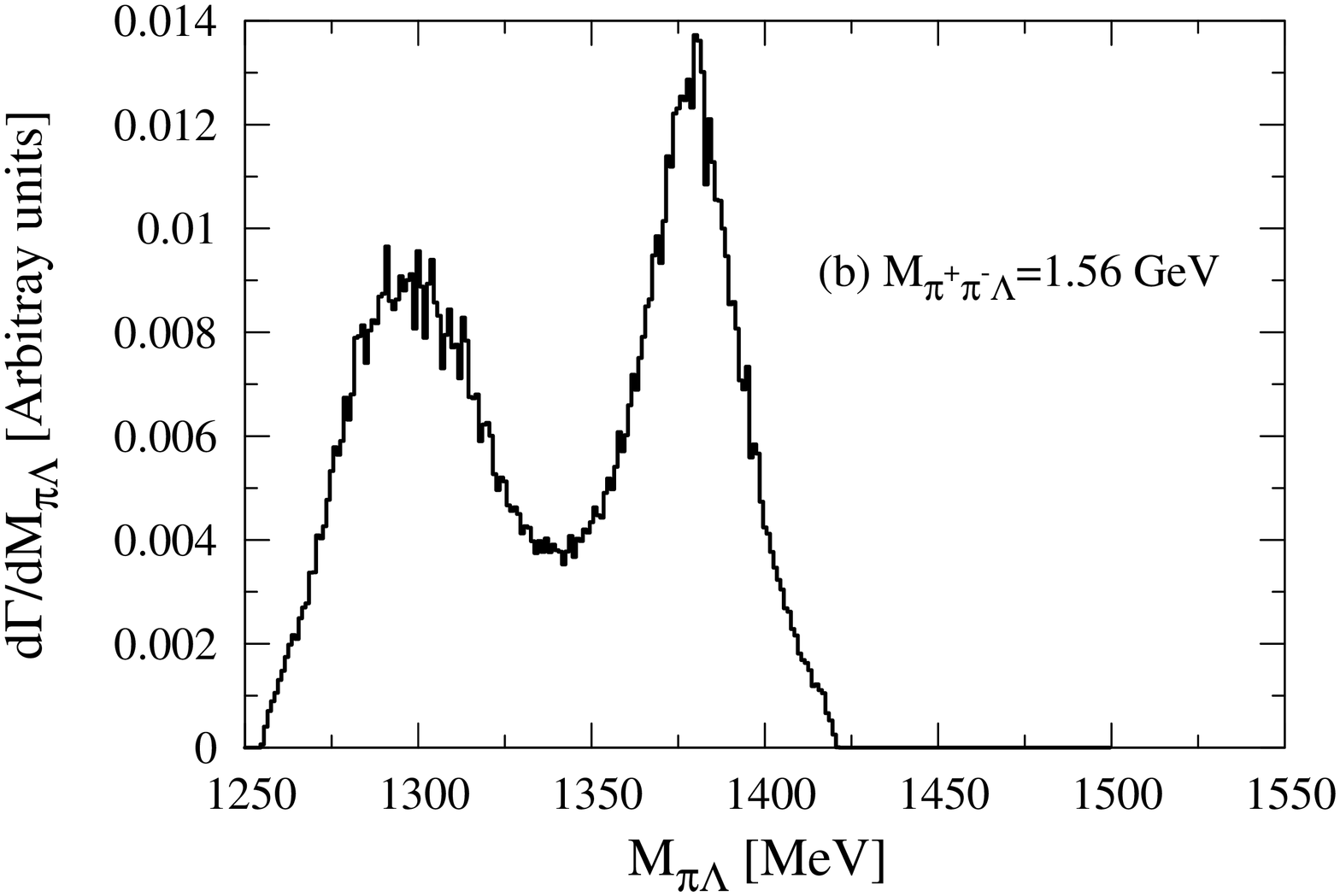}
\includegraphics[scale=0.35,angle=270]{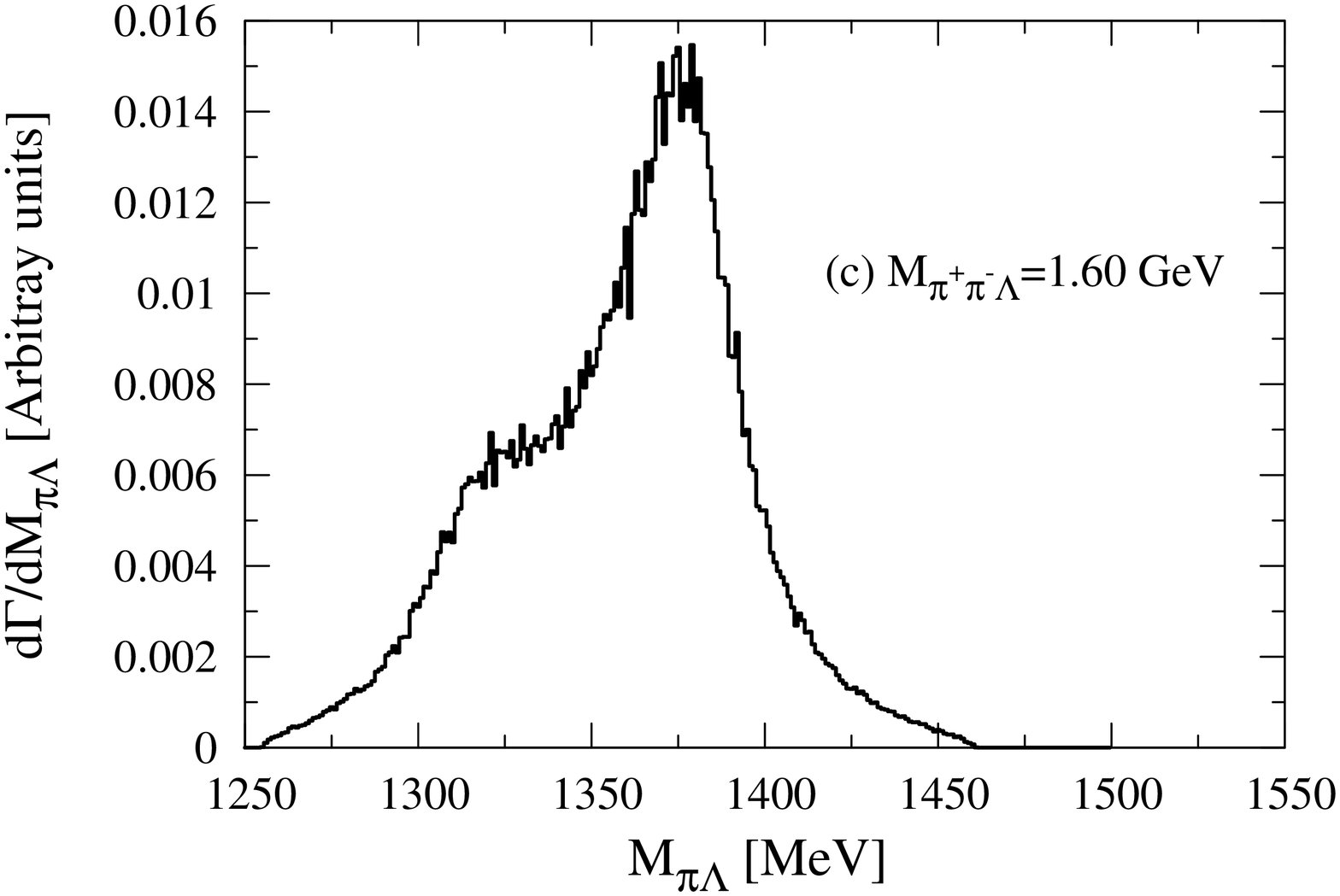}
\includegraphics[scale=0.35,angle=270]{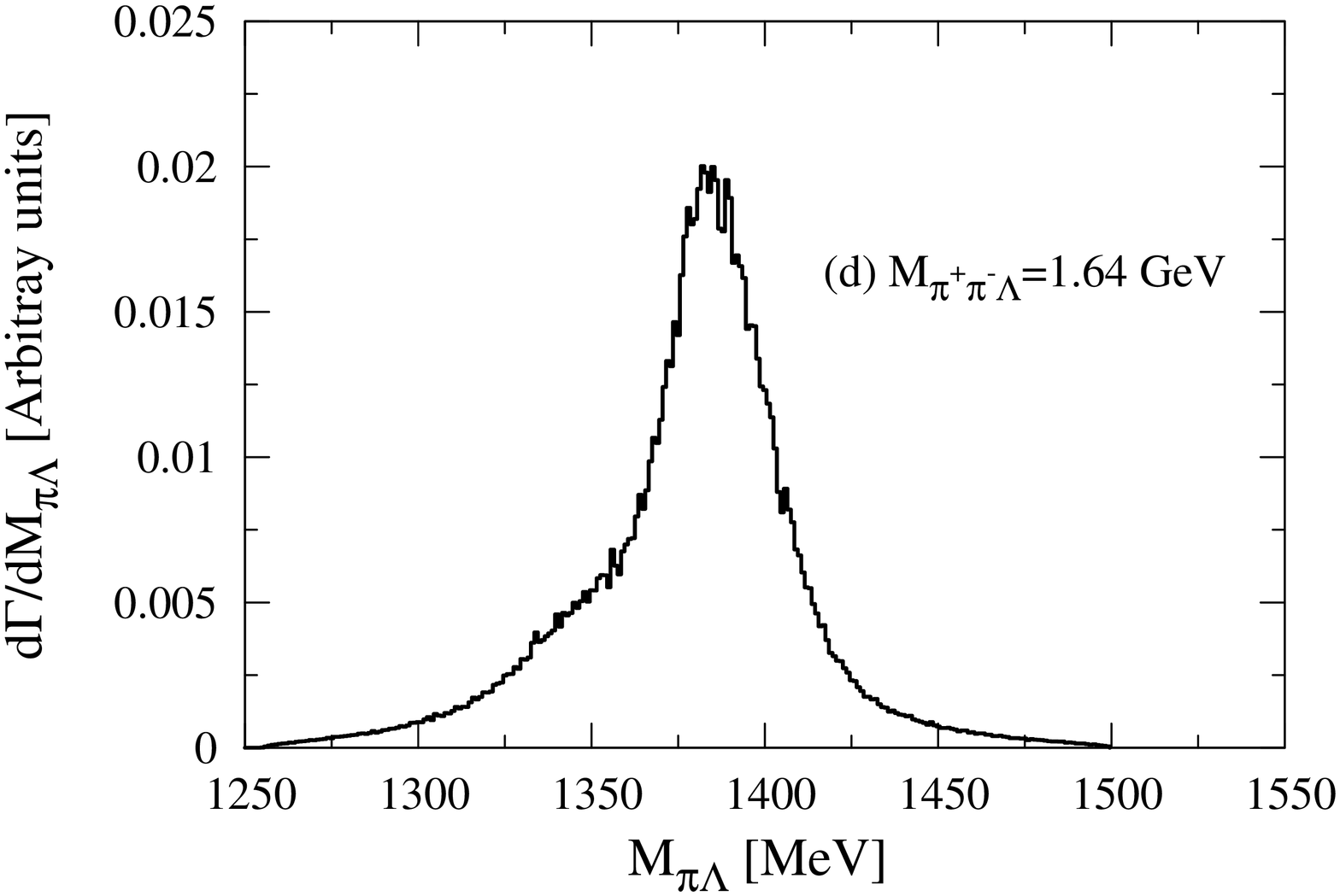}
\caption{The invariant mass distributions, $d\Gamma/dM_{\pi\Lambda}$, of the
$J/\psi\rightarrow\bar{\Lambda}\pi^+\pi^-\Lambda$ reaction at different $M_{\pi^+\pi^-\Lambda}$.
The curves have been normalized by their corresponding strength 
(some of them shown in Fig.~\ref{fig:ratio}) 
in such a way
that the areas under the curves are unity. \label{fig:sigmastar}}
\end{figure*}

In the chiral unitary picture, all the $\pi\pi\Lambda$ events
are generated through $\pi\Sigma^*$,  which should be  easily distinguishable experimentally. At the $\Lambda(1520)$ peak, one cannot
reconstruct the $\Sigma^*$ shape since there is no phase space for its production; yet, as one can see in Fig.~(\ref{fig:sigmastar}a), there
is a large concentration of strength at the upper threshold of $M_{\pi\Lambda}$ close to
the $\Sigma^*$ mass. Note that the peak at lower $\pi\Lambda$ invariant masses should be attributed
to the wrong $\pi\Lambda$ couples. By this we mean that if one looks at the $\pi^+\Lambda$ invariant mass,
one can see in Fig.~\ref{fig:ratio} that one produces $\pi^+\Sigma^{*-}$ together with 
$\pi^-\Sigma^{*+}$. Resonant $\pi^+\Lambda$ events come from the production of $\pi^-\Sigma^{*+}\rightarrow
\pi^-\pi^+\Lambda$, but when producing $\pi^+\Sigma^{*-}$, the $\pi^+\Lambda$ state is not resonant and it is what we call the ``wrong couple''. At higher $\pi\pi\Lambda$ invariant masses, however, the $\Sigma^*$ shape
shows up very clearly. 

In Fig.~(\ref{fig:sigmastar}b,c,d), we show the invariant mass distribution, $d\Gamma/dM_{\pi\Lambda}$, for other values of
$M_{\pi\pi\Lambda}$. In all these cases the $\Sigma^*$ peak in $M_{\pi\Lambda}$ shows up clearly and stays
at the same position independent of the $\pi\pi\Lambda$ mass. On the other hand, the ``background'' created by the wrong couple moves to higher $\pi\Lambda$ invariant masses as the $\pi\pi\Lambda$ mass grows, hiding below the $\Sigma^*$ peak at $M_{\pi\pi\Lambda}=1.64$ GeV, and moving to higher $M_{\pi\Lambda}$ values above
the $\Sigma^*$ peak as $M_{\pi\pi\Lambda}$ increases further (not shown). The strength for the $J/\psi\rightarrow\bar{\Lambda}
\pi^+\pi^-\Lambda$ reaction as a function of energy shown in Fig.~\ref{fig:ratio} in the vicinity of
the $\Lambda(1520)$ pole, together with the pattern of the $M_{\pi\Lambda}$ invariant masses shown in Fig.~\ref{fig:sigmastar}, are
predictions of the dynamics assumed into the mechanism of 
Fig.~\ref{fig:diagram}b, implying that all
$\pi\pi\Lambda$ production proceeds via $\pi\Sigma^*$ production, as
the chiral unitary model provides. The pattern of the $\pi\Lambda$
mass distribution at 1.52 GeV shown in Fig.~\ref{fig:sigmastar} should be a good reference 
for comparison with experiment. The predictions of the model at
$M_{\pi\Lambda}\geq 1.56$ GeV should be taken with more caution when comparing with experiment since this is a region where the background will not be negligible. Yet, the peak of the $\Sigma^*(1385)$
should be clearly seen.

The combined experimental study of these reactions and their mass distributions should definitely provide information to contrast the prediction of the chiral unitary approach as well as those of other
models, hence helping understand better the nature of the $\Lambda(1520)$.

Note that a similar
pattern of the $\pi\Lambda$ invariant masses as in Fig.~\ref{fig:sigmastar} was observed in the $K^-p\rightarrow \pi^0\pi^0\Lambda$
reaction~\cite{Prakhov:2004ri}. However, in that work the $K^-p$ total energy corresponds to values above
the $\Lambda(1520)$ mass and the data were not related to the properties of the 
$\Lambda(1520)$. In Ref.~\cite{Roca:2006sz} it was noticed that data for
$K^-p\rightarrow\pi^+\pi^-\Lambda$~\cite{Mast:1973gb} existed at the peak of the $\Lambda(1520)$, but not
at higher energies where the $\Sigma^*$ invariant mass could have been better reconstructed.
The work of  Ref.~\cite{Roca:2006sz} established the relationship between the two experiments,
linking them to the properties of the $\Lambda(1520)$.

\section{Summary}
We have studied the reactions $J/\psi\rightarrow\bar{\Lambda}K^-p$ and $J/\psi\rightarrow
\bar{\Lambda}\pi^+\pi^-\Lambda$, in particular, the ratio of their invariant mass distributions
at the $\Lambda(1520)$ peak, using
a unitary chiral approach. The unitary chiral approach contains chiral dynamics, which constrains 
the $s$-wave interaction, and parameters that are fitted to reproduce the $d$-wave amplitudes. It provides
a suitable framework to study processes where the $\Lambda(1520)$ plays a relevant role. In this
work, we concentrated on the ratio of these two invariant mass distributions around the $\Lambda(1520)$
peak position, which enabled us to overcome the unknown couplings of the $J/\psi$ to the $\bar{\Lambda}MB$
system, which are needed for the production mechanism.  The obtained invariant mass distributions
show clearly the $\Lambda(1520)$ peak, and the ratio at the peak position is $\sim4$. 

We also made predictions for the $\pi\Lambda$ mass distributions at different values of $M_{\pi\pi\Lambda}$,
close to the $\Lambda(1520)$ peak, which exhibit a peculiar pattern as a function of 
$M_{\pi\pi\Lambda}$ and show clearly the $\Sigma^*$ peak.  We encourage the
BES collaboration to perform an analysis of these two reactions. It will be extremely helpful
to improve the present experimental situation regarding the
$\Lambda(1520)$ resonance, particularly its $\pi\pi\Lambda$ decay mode, and 
to better understand the nature of the $\Lambda(1520)$ and its underlying chiral dynamics.

\section{Acknowledgments}
This work is partly
supported by FIS2006-03438, the National Natural Science Foundation
of China and the Chinese Academy of Sciences under
project number KJCX3-SYW-N2, the
Generalitat Valenciana and the EU Integrated
Infrastructure Initiative Hadron Physics Project under contract
RII3-CT-2004-506078.

\end{document}